\documentclass{my_aa}
\usepackage{natbib}
\usepackage{epsfig}
\usepackage{txfonts}

\begin{document}
\bibliographystyle{my_aa}
\title{A transiting planet among 23 new near-threshold candidates from the OGLE survey\thanks{Based on observations made with the FORS1 camera and the FLAMES/UVES spectrograph at the VLT, ESO, Chile (programmes 07.C-0706 and 177.C-0666) and 1.3-m Warsaw Telescope at Las Campanas
Observatory, Chile.} | OGLE-TR-182}
\author{F. Pont$^1$, O. Tamuz $^2$, A. Udalski$^{3,4}$, T. Mazeh $^2$, F. Bouchy $^5$,  C. Melo$^{9}$,  D. Naef$^{9}$, C. Moutou$^{10}$, R. Diaz$^{13}$,  W.  Gieren$^{7}$,  M. Gillon$^{1}$, S. Hoyer$^{11}$, M. Kubiak$^{3,4}$, M. Mayor$^{1}$, D. Minniti$^{8}$,  G. Pietrzynski$^{3,4,7}$, D. Queloz$^1$, S. Ramirez$^{8}$, M. T. Ruiz$^{11}$, N.C. Santos$^{12}$,  I. Soszy{\'n}ski$^{3,4}$, O. Szewczyk$^{3,4}$, M.K. Szyma{\'n}ski$^{3,4}$, S. Udry$^1$, K. Ulaczyk$^{3,4}$, {\L}. Wyrzykowski$^{4,6}$, M. Zoccali$^{8}$}
\offprints{frederic.pont@obs.unige.ch}
\institute{$^1$  Observatoire de Gen\`eve, 51 Chemin des Maillettes,1290 Sauverny,
    Switzerland\\
 $^2$ School of Physics and Astronomy, R. and B. Sackler Faculty of Exact Sciences, Tel Aviv University, Tel Aviv, Israel\\   
  $^{3}$ Warsaw University Observatory, Al. Ujazdowskie 4, 00-478, Warsaw, Poland\\
  $^4$ The OGLE Team\\
$^5$ Institut d'Astrophysique de Paris, 98bis Bd Arago, 75014 Paris, France\\
$^6$ Institute of Astronomy, University of Cambridge, Madingley Road,
Cambridge CB3 0HA, UK\\
$^7$ Departamento de Fisica, Astronomy Group, Universidad de Concepci\'on, Casilla 160-C, Concepci\'on, Chile\\
$^8$ Departmento de Astronom\'ia y Astrof\'isica, Pontificia Universidad Cat\'olica de Chile, Casilla 306, Santiago 22, Chile\\
$^9$ European Southern Observatory, Casilla 19001, Santiago 19, Chile\\
$^{10}$ Laboratoire d'Astrophysique de Marseille, Traverse du Siphon, BP8, Les Trois Lucs, 13376 Marseille
  cedex 12, France\\    
$^{11}$ Department of Astronomy, Universidad de Chile, Santiago, Chile\\
$^{12}$ Centro de Astrof\'\i sica, Universidade do Porto, Rua das Estrelas, 4150-762 Porto, Portugal\\
$^{13}$ Instituto de Astronomia y F\'\i sica del Espacio, Buenos Aires, Argentina }

\date{Received date / accepted date}

   \authorrunning{F. Pont et al.}
   \titlerunning{OGLE-TR-182 -- a new transiting planet among 23 near-threshold candidates}
\abstract{By re-processing the data of the second season of the OGLE survey for planetary transits and adding new mesurements on the same fields gathered in subsequent years with the OGLE telescope, we have identified 23 new transit candidates, recorded as OGLE-TR-178 to OGLE-TR-200. We studied the nature of these objects with the FLAMES/UVES multi-fiber spectrograph on the VLT. One of the candidates, OGLE-TR-182, was confirmed as a transiting gas giant planet on a 4-day orbit. We characterised it with further observations using the FORS1 camera and UVES spectrograph on the VLT. OGLE-TR-182b is a typical ``hot Jupiter'' with an orbital period of 3.98 days, a mass of 1.01 $\pm 0.15$ M$_{Jup}$ and a radius of 1.13 $^{+0.24}_{-0.08}$ R$_{Jup}$.  Confirming this transiting planet required a large investment in telescope time with the best instruments available, and we comment on the difficulty of the confirmation process for transiting planets in the OGLE survey. We delienate the zone were confirmation is difficult or impossible, and discuss the implications for the Corot space mission in its quest for transiting telluric planets.  

}

\maketitle

\section{Introduction}

Transiting extrasolar planets are essential to our understanding of planetary structure, formation and evolution outside the Solar System. The observation of transits and secondary eclipses gives access to such quantities as a planet's true mass, radius, density, surface temperature and atmospheric spectrum. The first transiting exoplanet was identified in 1999 around HD 209458. In the past three years, transiting exoplanet have been found in rapidly increasing number, both by radial velocity planet searches and by photometric surveys\footnote{For an updated list see obswww.unige.ch/$\sim$pont/TRANSITS.htm}. The OGLE search for transiting planets and low-mass stellar companions \citep{ud02a} has been the first photometric transit survey to yield results. The first three seasons of photometric observations have revealed 137 transit candidates \citep{ud02a,ud02c,ud02b,ud03}, among which 5 planets were found \citep{ko03,bo04,po04a,kon04,bo05,ko05}, as well as two planet-sized low-mass stars \citep{po05b,po06b}. Three further seasons of the OGLE transit survey have now been completed and await publication (Minniti et al., Udalski et al., in prep.).

The spectroscopic follow-up of most of the 137 first OGLE transit candidates, presented in \citet{bo05} and \citet{po05a}, has shown that the vast majority of the transit candidates were eclipsing binaries. A rate of one transiting planet for 10-20 eclipsing binaries is typical. A higher rate of planets can be found among candidates near the detection threshold. Two of the five planets from the OGLE survey, OGLE-TR-56 and OGLE-TR-132, were identified as candidates only after the application of a more sensitive transit detection algorithm \citep{ko02}. However, lowering the detection threshold comes at the price of including some false positives of the detection procedure.
The objective of the present study is to explore the regime near the detection threshold, the zone where the ratio of planets to eclipsing binaries will be more favourable than for deeper transit signals, but where the reality of the signal itself is not beyond doubt.  The exploration of this zone is relevant not only to identify new transiting planets in the OGLE survey, but also because other transit surveys will face similar issues, notably the {\it CoRoT} and {\it Kepler} space-based transit searches.

\section{Candidate selection}

The observations described in \citet{ud02c} were pooled with more recent data obtained on the same field with the OGLE telescope. The data consists of 1200-1400 measurements of $\sim 10^5$ stars, spread over 3 years, on $1.25 \times 1.25$ degree fields in the Carina section of the Galactic plane.

\citet{po06} have examined the behaviour of the detection threshold in ground-based photometric transit surveys, with a closer focus on the OGLE survey, and shown how the presence of unaccounted trends and systematics in the photometric data define the detection threshold and can impede the detection of most transiting planets in the sample. Several schemes have been devised to remove trends of unknown origin in transit-search photometric times series, including  the "Trend filtering algorithm" of  \citet{kov05}, and the "Sysrem" algorithm of \citet{ta05}. The principle of these algorithm is that the dataset is examined as a whole for systematic effects that affect all lightcurves in a similar manner, modulo different coefficients for each object. The Sysrem algorithm calls the effects ``generalized airmasses'' and the coefficients ``generalised colours''. In the same way that an {\it airmass $\times$ colour} term is fitted to each lightcurve to remove differential refraction effects in photometry, the algorithm finds multiplicative effects (whose origin need not be known) affecting each lightcurve differently.

We applied the ``Sysrem'' algorithm to the OGLE data, then ran an updated version of the BLS transit-search algoritm \citet{ko02}. We examined the most significant candidates identified, and built a list of 23 candidates for follow-up.  We attempted to place the selection threshold low enough so as to reach the level were false positives and real transits are found in comparable numbers. 

The relevant characteristics of the 23 candidates are listed in Table~\ref{candidates}. They are named according to the usual convention of the OGLE transit survey.  The finding charts and lightcurves are available on the OGLE website.

\begin{table*}[ht!]
\begin{tabular}{l l l r l l r r r}
\hline
Name & Coordinates &$I$ &	Period & Epoch & Depth 	&  SDE	 &$N_{tr}$	 &$n_{tr}$	 \\
   &  [2000] & [mag]  & [days] & [JD-2450000] & &  & & \\ \hline
OGLE-TR-178 & 11:07:35.25 \ $-$61:21:35.9 & 16.56&2.97115&2547.66173&0.016&26.05&19&134\\
OGLE-TR-179 & 11:09:10.99 \ $-$61:21:44.4 & 15.13&12.67106&2554.79150&0.034&20.19&4&26\\
OGLE-TR-180 & 11:07:15.36 \ $-$61:16:02.7 & 16.74&1.99601&2546.29696&0.012&17.75&34&105\\
OGLE-TR-181 & 11:09:26.34 \ $-$61:08:21.8 & 16.29&2.38960&2550.18511&0.010&11.09&44&10\\
OGLE-TR-182 & 11:09:18.84 \ $-$61:05:42.8 & 15.86&3.98105&2551.70430&0.010&17.68&10&79\\
OGLE-TR-183 & 11:07:05.34 \ $-$61:01:07.1 & 15.32&4.78217&2543.17799&0.015&10.28&2&15\\
OGLE-TR-184 & 11:07:24.29 \ $-$60:58:03.7 & 15.57&4.92005&2549.73531&0.015&12.10&12&45\\
OGLE-TR-185 & 11:07:30.06 \ $-$60:53:12.4 & 16.72&2.78427&2547.66899&0.035&12.43&8&27\\
OGLE-TR-186 & 11:08:12.56 \ $-$60:51:16.9 & 16.54&14.81481&2559.90618&0.054&17.10&3&19\\
OGLE-TR-187 & 11:06:17.33 \ $-$60:51:11.7 & 14.07&3.45686&2554.41237&0.008&10.13&6&15\\
OGLE-TR-188 & 11:06:23.98 \ $-$60:56:16.7 & 16.38&6.87663&2554.10247&0.031&11.43&5&16\\
OGLE-TR-189 & 11:04:40.23 \ $-$61:21:57.4 & 15.03&1.73937&2549.22368&0.006&13.59&21&120\\
OGLE-TR-190 & 11:06:18.65 \ $-$61:16:18.8 & 16.06&9.38262&2549.20966&0.043&10.54&4&17\\
OGLE-TR-191 & 10:57:44.85 \ $-$61:49:20.2 & 15.57&2.51946&2561.57557&0.007&10.23&11&40\\
OGLE-TR-192 & 10:57:35.48 \ $-$61:34:30.5 & 14.41&5.42388&2557.36624&0.008&7.69&6&16\\
OGLE-TR-193 & 10:59:33.97 \ $-$61:23:16.7 & 14.99&2.95081&2557.80469&0.008&13.47&14&41\\
OGLE-TR-194 & 10:55:50.15 \ $-$61:35:37.5 & 14.69&1.59492&2557.15087&0.006&17.17&23&92\\
OGLE-TR-195 & 10:56:41.30 \ $-$61:32:06.2 & 14.19&3.62174&2557.71767&0.006&10.00&11&54\\
OGLE-TR-196 & 10:56:15.91 \ $-$61:51:30.9 & 15.57&2.15540&2557.61541&0.012&12.03&11&40\\
OGLE-TR-197 & 10:54:47.17 \ $-$61:22:03.9 & 14.59&2.40587&2607.54146&0.019&19.36&14&68\\
OGLE-TR-198 & 10:52:07.33 \ $-$61:22:07.1 & 15.44&13.63141&2616.90713&0.018&13.78&4&33\\
OGLE-TR-199 & 10:50:32.77 \ $-$61:35:17.1 & 14.88&8.83470&2603.09783&0.017&13.65&3&19\\
OGLE-TR-200 & 10:50:56.51 \ $-$61:55:53.3 & 15.63&6.48845&2606.06723&0.023&16.96&4&25\\ \hline

\end{tabular}
\caption{Planetary and low-mass star transit candidates. {\it SDE} is the significance indicator of the \citet{ko02} transit detection algorithm -- the signal-to-noise ratio of the transit detection. $N_{tr}$ is the number of transits covered by the photometric measurements, and $n_{tr}$ the number of data points in the transit.}
\label{candidates}
\end{table*}

\section{FLAMES observations and results}

Spectroscopic observations of the candidates were obtained during four half-nights with the FLAMES multi-fiber spectrograph in 24-28 February 2006 (ESO 07.C-0706). One candidate, OGLE-TR-182, turned out to be especially interesting and was followed during the ESO Large programme on OGLE transits (ESO 177.C-0666). It is considered in more details in Section~\ref{tr182}. 

The FLAMES spectrograph has been used to detect or characterise the five previous planets from the OGLE survey. The relevant details, as well as the methods to sort out transiting planets from eclipsing binaries and other types of false positives, can be found in \citet{bo05} and \citet{po05a}.

Since we had very poor weather throughout our VLT run, we resorted to a ``fast track'' approach of dropping candidates as soon as it was clear that they were not detectable planets, before their actual nature was solved. After one spectroscopic measurement, candidates with broadened spectral lines, shallow lines or double-lined spectra were dismissed (probable eclipsing binaries or blends). After two spectra, candidates varying by more than a few km$\,$s$^{-1}$ were also dismissed. Therefore, unlike for previous seasons, we did not resolve the nature of the non-planetary candidates. In particular, we did not determine the ratio of eclipsing binaries to pure false positives in our sample. 

Nine candidates were not observed.   Three because they were fainter than $I=16.5$. Prohibitive observing time is needed to confirm a planet at such magnitudes. Six because they were situated at isolated locations on the sky, requiring a FLAMES setup to observe only one object, and they were not among the highest-priority candidates. 

Of the fourteen remaining candidates, only one candidate survived the initial screening. It was monitored in 2006 and 2007, with FLAMES in radial velocity, in photometry with the FORS camera, and in spectroscopy with UVES in slit mode (see Section~\ref{tr182}). The status of the other candidates after the spectroscopic follow-up is given in Table~\ref{nature}.

\begin{table}
\centering
\begin{tabular}{l  l}
\hline
Name		& Results \\ \hline
OGLE-TR-178	 			&  	faint target, not observed 	\\
OGLE-TR-179		&	flat CCF\\
OGLE-TR-180	 			&		faint target, not observed\\
OGLE-TR-181	 	& fast rotator (synch.?)\\
OGLE-TR-182			& transiting planet\\
OGLE-TR-183	 		& fast rotator (synch.?)\\
OGLE-TR-184	 		& fast rotator (synch.?)\\
OGLE-TR-185	 		& fast rotator (synch.?)\\
OGLE-TR-186	 			& faint target, not observed\\
OGLE-TR-187	 		& double-lined spectroscopic binary\\
OGLE-TR-188			& blend of two line systems\\
OGLE-TR-189	 			 &not observed\\
OGLE-TR-190	 			&not observed	\\
OGLE-TR-191		&	fast rotator (synch.?)\\
OGLE-TR-192	  	&	flat CCF\\
OGLE-TR-193	 		&not observed\\
OGLE-TR-194	 		&flat CCF\\
OGLE-TR-195	 		&not obseved\\
OGLE-TR-196	 		&fast rotator (synch.?)\\
OGLE-TR-197			&flat CCF\\
OGLE-TR-198	 			&not observed	\\
OGLE-TR-199			&single-lined spectroscopic binary\\
OGLE-TR-200	 			&not observed\\ \hline
\end{tabular}

\caption{Results of the spectroscopic follow-up. {\it Synch.?}: possible synchronous rotation with the period of the transit signal, indicating an eclipsing binary. {\it Flat CCF}: no signal in the cross-correlation function, indicating a fast-rotating star or early-type star.}
\label{nature}
\end{table}

\section{Analysis of OGLE-TR-182}
\label{tr182}

\begin{figure}
\resizebox{8cm}{!}{\includegraphics{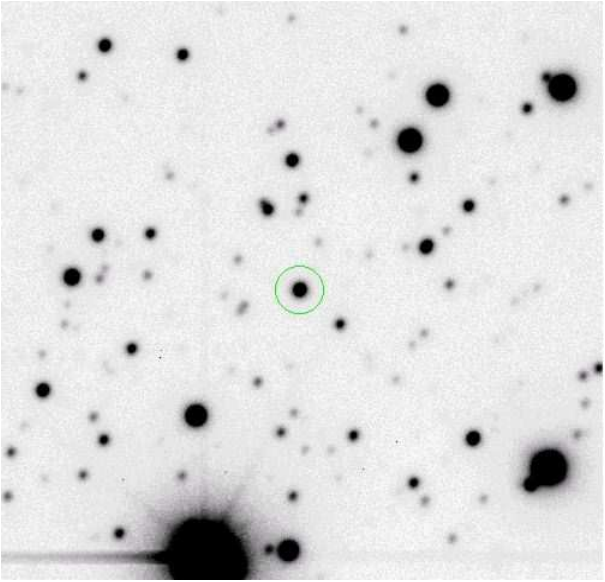}}
\caption{Finding chart for OGLE-TR-182 in the I-band acquired with the VLT.
The field is 50'x50', North is up and East to the left. The star
does not show any nearby neighbours $>$5 magnitudes fainter.}
\label{fc182}
\end{figure}

The planetary transit candidate OGLE-TR-182 is an $I=15.86$ magnitude star in Carina. Its coordinates are given in Table~\ref{candidates} and a finding chart is shown on Fig.~\ref{fc182}. 

\subsection{Photometric observations}

The preliminary photometric ephemeris of OGLE-TR-182 was based on a
series of transits observed during the original OGLE run carried
out in the 2002 observing season. Because the orbital period is very close
to an integer number of days, transits observed from a given
geographical location occur in observing windows separated by long gaps
when only non-transit phases are available for observations. No trace of
additional transits were found in the OGLE data from the remaining
observing seasons.

To refine the ephemeris and confirm that the candidate transit signal from 2002 
 was not a false positive, an extensive hunt for additional
transits was carried out by OGLE in the 2007 observing season. Although
the transit signal was not caught again, the collected photometry together with
spectroscopic timing allowed to strictly constrain the possible transit
occurrence. The VLT photometric time allocated in May and June 2007 to
our project was used to catch the new series of transits of OGLE-TR-182
and derive a much more precise transit shape. With five years 
of baseline,  the derived photometric ephemeris is now very secure.

\begin{figure}
\resizebox{8cm}{!}{\includegraphics{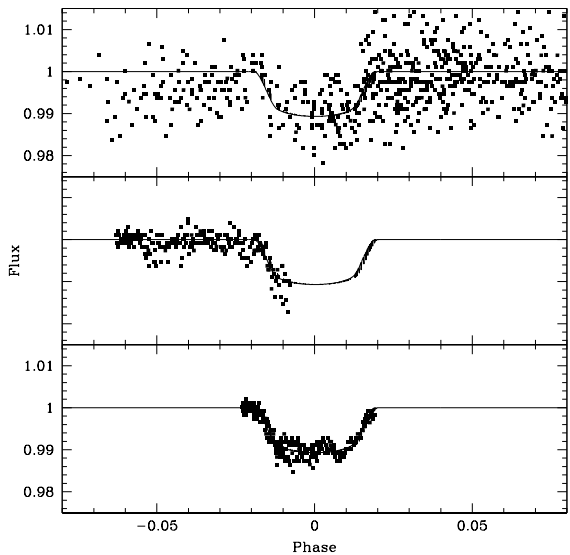}}
\caption{Lightcurve of OGLE-TR-182 with the OGLE telescope (top) and the FORS1 camera on the VLT on June 6 (middle) and June 18 (bottom) 2007.}
\label{phot}
\end{figure}

Complete coverage of the transit of OGLE-TR-182 was obtained with the FORS camera on the VLT, in the nights of May 22, June 6 and June 18, 2007. The observation strategy was identical to that used for OGLE-TR-10 and OGLE-TR-56 in \citet{po07}.  The time series are shown in the middle and bottom panels of Figure~\ref{phot}. Exposure were taken every minute and the typical photon noise is 2 mmag. The reduction was carried out with the OGLE pipeline. The data are available on request to the authors. 

\subsection{Radial-velocity observations}

The target OGLE-TR-182 was observed 20 times with FLAMES/UVES. The resuling radial velocity measurements are given in Table~\ref{vr}. A periodic variation is found in radial velocity of OGLE-TR-182 with a period (P$\simeq$ 3.979 days) and phasing compatible with the photometric signal. Because of the period very close to 4 days, measurements had to be spread over two seasons to cover a sufficient part of the phase.

From an analysis of the FLAMES measurements over all the objects followed during our runs, we find that systematics zero-point shifts with r.m.s. 40--60 m s$^{-1}$ need to be added in quadrature to the photon-noise radial velocity uncertainties to account for the observed residuals. This is higher than in our 2004 previous run \citet{bo04}. We attribute this to the very poor weather conditions in most of our runs and to the fact that the data acquisition was spread over different runs separated by several months, with some possible contribution from stellar activity.

The evolution of the spectral lines shape was examined to rule out blend scenarios. No line bisector variation correlated with radial velocity or orbital phase was observed.

\begin{table}
\begin{tabular}{l l l}
\hline
Date & VR & $\sigma_{VR}$  \\
$$[JD-2450000] & [km$,$s$^{-1}$] & [km$\,$s$^{-1}$] \\ \hline
3791.744140  &  22.413  &  0.040  \\
3793.662311  &  22.250  &  0.112  \\
3793.763507  &  22.212  &  0.054  \\
3794.713268  &  22.339  &  0.040  \\
3794.873518  &  22.252  &  0.043  \\
3800.801722  &  22.468  &  0.039  \\
3852.540037  &  22.317  &  0.048  \\
3853.575399  &  22.189  &  0.041  \\
3854.499952  &  22.295  &  0.040  \\
3856.619260  &  22.451  &  0.045  \\
3858.616679  &  22.406  &  0.041  \\
3859.727660  &  22.461  &  0.058  \\
4141.610000  &  22.400  &  0.030  \\
4143.780490  &  22.191  &  0.047  \\
4144.743801  &  22.325  &  0.047  \\
4144.773062  &  22.256  &  0.041  \\
4145.774175  &  22.504  &  0.072  \\
4148.651117  &  22.330  &  0.045  \\
4149.777455  &  22.454  &  0.050  \\
4150.739605  &  22.452  &  0.041  \\ \hline
\end{tabular}
\caption{Radial velocity measurements for OGLE-TR-182.}
\label{vr}
\end{table}

\begin{figure}
\resizebox{8cm}{!}{\includegraphics[angle=-90]{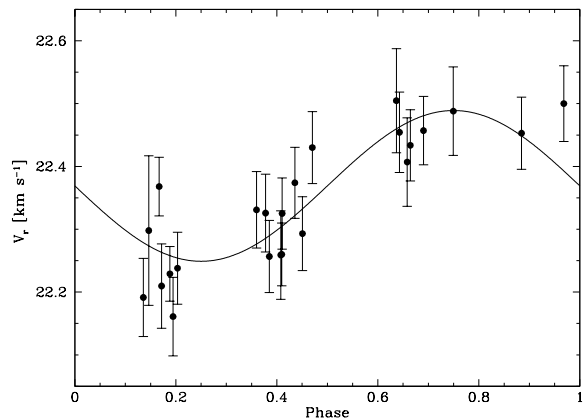}}
\caption{Radial velocity observations for OGLE-TR-182, phased with the photometric transit signal.}
\label{spec}
\end{figure}

\subsection{Spectroscopic observations}

OGLE-TR-182 was observed for a total of 7 hours in service operation in the summer of 2007 with UVES in slit mode, in order to acquire a high signal-to-noise spectrum for the determination of atmospheric parameters (total S/N $\sim$ 90). The UVES spectra are not suitable for precise radial velocity measurements, because in slit mode centering offsets on the sky translate to large velocity zero-point changes.

The reduction strategy and data analysis were identical to that in \citet{san06}. The resulting temperature, metallicity and gravity are given in Table~\ref{spectro}.

\subsection{Transit analysis}

A transiting planet signal was fitted to the photometry and radial velocity, asssuming a null orbital eccentricity. No sign of non-zero eccentricity is observed in the radial velocity data or in the timing of the transits compared to the radial velocity orbit. Since nearly all short-period planets have been circularized by tidal interactions, it is justified to assume a circular orbit unless there are clear indications of the contrary. We used the \citet{ma02} description of transit profiles with the \citet{cl00} limb-darkening coefficients. We determined the uncertainties accounting for the photometric red noise as prescribed in \citet{po06}. 
Since no complete transit with a sufficient out-of-transit baseline was observed, we derived the estimates of the stellar mass and radius solely from the comparison of the spectroscopic parameters with  \citet{gi02} stellar evolution models by maximum-likelihood. 
The resulting parameters for the system are given in Table~\ref{param}. 

\begin{table}[th] 

\begin{tabular}{l l}
\hline
Period       [days]      & 3.97910 $\pm$ 0.00001\\
Transit epoch [JD] & 2454270.572 $\pm$ 0.002 \\
VR semi-amplitude  [m/s] &  120 $\pm$ 17\\
Semi-major axis [AU]& 0.051 $\pm$ 0.001  \\
Radius ratio & 0.102 $\pm$ 0.004   \\
Orbital angle [$^0$] & 85.7 $\pm$ 0.3  \\
 &   \\

$T_{eff}$ [K]& 5924 $\pm$ 64  \\
$\log g$ & 4.47$\pm$0.18 \\
$\eta$ [km$\,$s$^{-1}$]& 0.91$\pm$0.09 \\
$$[Fe/H] & 0.37+-0.08 \\ 
Star radius  [$R_\odot$]&  1.14 $^{+0.23}_{-0.06}$ \\
Star mass [$M_\odot$]&  1.14 $\pm$ 0.05 \\
 &   \\
Planet radius [$R_J$]&   1.13 $^{+0.24}_{-0.08}$\\
Planet mass [$M_J$]&   1.01 $\pm$ 0.15\\ \hline
\end{tabular}

\caption{Parameters for the OGLE-TR-182 system}
\label{param}
\label{spectro}
\end{table}

\section{Discussion}

\subsection{OGLE-TR-182b as a transiting planet}

The companion of OGLE-TR-182 has parameters typical of the planets detected by photometric transit surveys in all respects: it orbits a high-metallicity dwarf star, it has a mass comparable to that of Jupiter and a slightly larger size. Its period is close to an integer number of days, reflecting the strong selection bias due to the window function \citep[see e.g.][about OGLE-TR-111$b$, another $P\sim 4$ days transiting planet]{po04a}. At present, the constraint on the planetary radius is not sufficient to determine whether its radius corresponds to model expectations or whether is belongs to the set of anomalously large transiting hot Jupiters. Its position in the mass-period diagram is also similar to other known transiting planets, and reinforces the link between mass and period for close-in gas giants first pointed out by \citet{ma05}.

\subsection{The ``Twilight Zone'' of transit surveys}

\begin{figure*}[ht]
\resizebox{16cm}{!}{\includegraphics[angle=-90]{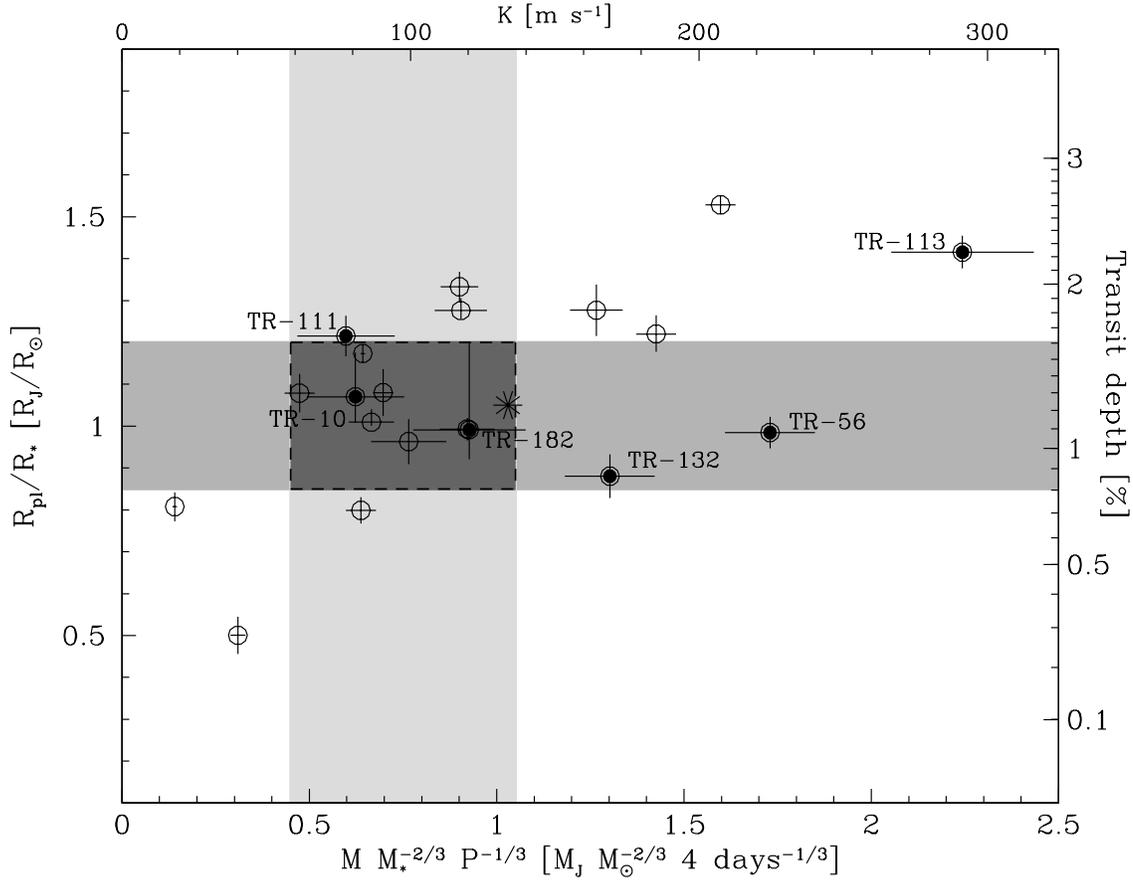}}
\caption{''Observational'' mass-radius plot for known transiting exoplanets. The horizontal axis is the planet mass, scaled to $M^{2/3}P^{1/3}$, proportional to the radial-velocity semi-amplitude $K$. The vertical axis is the planet radius scaled to the host star radius, related to the depth of the photometric transit signal. "TR-" labels refer to OGLE candidates (closed symbols). Open symbols mark the position of other known transiting planets. The star symbol marks the best-fit location of another unsolved planet candidate from the third OGLE season. The gray bands show the near-threshold zones for the photometric detection (horizontal) and the spectroscopic confirmation (vertical) in the case of the OGLE survey. The dashed area is the ``twilight zone'' defined in the text, where confirmation is problematic.}
\label{mr}
\end{figure*}

The confirmation follow-up process for OGLE-TR-182 necessitated more than ten hours of FLAMES/VLT time for the radial velocity orbit, plus a comparable amount of FORS/VLT time for the transit lightcurve. In addition, several unsuccessful attempts were made to recover the transit timing in 2007 with the OGLE telescope, and 7 hours of UVES/VLT were devoted to measuring the spectroscopic parameters of the primary. This represents a very large amount of observational resources, and can be considered near the upper limit of what can be reasonably invested to identify a transiting planet. 

Therefore, OGLE-TR-182 is a useful object to quantify the zone were neither the photometric signal nor the radial velocity signal are clear beyond doubt, the ``twilight zone'' of planetary transit candidate confirmation. In these cases, confirming the nature of the system is very difficult and time consuming. When the photometric signal is a possible false positive, a clear radial velocity orbit at the same period is an essential confirmation | as was for instance the case for OGLE-TR-132 \citep{bo04}. On the other hand, when the radial velocity signal is marginal, a clear transit signal allows the phase and period of the orbit to be determined with confidence, reducing the radial-velocity orbit fit to a two-parameter problem ($V_0$, the systemic velocity,  and $K$, the orbital semi-amplitude) | as was the case for OGLE-TR-10 \citep{bo05,ko05}. 

However, when both the photometric and spectroscopic signals are marginal, many more observations are necessary until reasonable certainty can be achieved about the presence of a planetary companion. The uncertainties on the lightcurve make it difficult to phase the radial velocity data. The high radial velocity uncertainties hinder the identification of an orbital motion with the correct period, and the elimination of eclipsing binary blend scenarios

The OGLE survey is the first to explore this ``twilight zone'' in real conditions, since other ground-based surveys target brighter stars, for which very precise radial velocities can be obtained, so that the significance of the radial velocity signal can be established relatively easily \citep[e.g.][]{ca07}. Based on the cases of OGLE-TR-10, OGLE-TR-132 and OGLE-TR-182,  and on the discussion in \citet{po06}, we define the limits of the follow-up twilight zone as follows:

\noindent
-- photometric transit detection with $8< S_r < 12$, where $S_r$ is the transit significance in the presence of red noise \citep[see definition in][]{po06} 

\noindent
-- radial velocity orbital semi-amplitude 1-2 times the radial velocity uncertainties for 1-hour exposures with the facilities available: $\sigma_{vr} < K < 2\cdot \sigma_{vr} $

These limits can be translated, for a circular orbit and a central transit, into limits on the radius and mass of the planet. Figure~\ref{mr} shows the ``observational'' mass-radius diagram for the known transiting exoplanets. The horizontal axis is the planet mass divided by $M_*^{2/3}P^{1/3} $, to make it proportional to the observed radial velocity semi-amplitude ("$K$"). The vertical axis is the planet radius divided by the radius of the star, to make it proportional to the squareroot of the transit depth (at wavelengths where stellar limb darkening can be neglected). The units are such that a Jupiter-sized planet transiting a solar-sized star on a 4-day orbit will be placed at (1;1). Objects at similar positions in this plot will present similar challenges for confirmation. The \mbox{ ``TR-''} labels refer to OGLE candidates, the other unmarked points are transiting planets from other surveys. The gray horizontal band is the zone where the transit detection is near the detection threshold, the vertical band is the zone where the radial velocity orbital signal is near the threshold.
The intersection of the two,  delimited by the dashed lines, represents the ``twilight zone'' for the OGLE survey. We use a red noise level of $\sigma_r = 3$ mmag  \citep{po06}, a radial velocity uncertainty of 60-70 m/s (photon noise plus systematics), and assume that 5-10 transits are observed by the photometry. We find $ 0.85  < R_{pl}/R_* < 1.20$ and $ 0.45 < M_{pl}/{M_*}^{-2/3} P^{-1/3} < 1.05$ [$M_J {M_\odot}^{-2/3} ({4\ {\rm days}})^{-1/3}$] for the zone boundaries. Candidates in this zone will be very difficult or impossible to confirm. On the left of the zone, radial velocity confirmation is out of reach, and below, the transit signal is below the photometric detection threshold.

The twilight zone for the OGLE transit survey encompasses the region in the mass-radius diagram corresponding to a normal hot Jupiter around a solar-type star. Hence the difficulty of the OGLE survey to detect transiting gas giants unless they are exceptionally heavy with a very short period like OGLE-TR-56, OGLE-TR-113 and OGLE-TR-132, or have an exceptionally high radius ratio like OGLE-TR-111.

The planetary transit system OGLE-TR-10 is also located within the zone, and indeed confirmation of its planetary nature required large investments in follow-up means both with the VLT \citep{bo05} and Keck  \citep{ko05} 8-10 meter telescopes. 

As a further illustration of the extent of the zone, the plot shows the position of another candidate from the OGLE survey, yet unsolved despite extensive measurements with FLAMES and FORS. The radial velocity data is compatible with a planetary orbit, but many more observations would be required to confirm it securely. The best-fit planetary solution for this object is plotted on Fig.~\ref{mr}.

In the wider context of transit searches in general, it is interesting to find where the ``twilight zone'' is located for different surveys, especially the space-based transit searches {\it CoRoT} and {\it Kepler}. For wide-field, small-camera surveys like HAT, WASP, XO and TrES, the twilight zone is not an important issue. Because the candidates are brighter, standard planet-search spectrographs can be used for the radial velocity follow-up, and the zone moves to the left of the mass-radius diagram, in a region were no planets are expected (low-mass, Jupiter-sized planets). In other words, if a planet is large enough to be detected by these surveys, it produces a radial velocity signal that is easily picked up by Doppler spectrographs.

Deeper surveys like SWEEPS \citep{sa06} and planet searches in star clusters also have no twilight zone problem, for the opposite reason: their candidates are too faint to be confirmed in radial velocity for Jupiter-like planetary masses. 

In the case of the {\it CoRoT} space transit search, the zone will be located in a key position. In planet radius, it is expected to cover the 2-4 $R_\oplus$ range \citep[see][]{mo05} for the brightest targets. In planet mass, using the HARPS spectrograph, it will be in the 5-20 $M_\oplus $ range, for short periods. This is a zone were planets are thought to be numerous, the domain of the ``hot Neptunes'' and ``super-Earths''. Indeed, the detection of this type of planets constitutes the main objective of the {\it CoRoT} planetary transit search. From our experience with the OGLE follow-up, we therefore conclude that the {\it CoRoT} mission will face similarly difficult cases in the confirmation process of transiting planets. The telescope time necessary for the follow-up of these candidates should be adequately evaluated. The OGLE follow-up process can provide some useful guidelines.

\begin{acknowledgements}
Part of the spectroscopic and photometric observations were obtained in the context of the `666 collaboration on OGLE transits'.  The authors wish to thank David Weldrake and Artur Narwid for obtaining
additional photometry for OGLE-182 with the Australian National University
40-inch (1m) telescope, which helped constrain the planetary orbit. The OGLE project is partially supported by the Polish MNiSW grant N20303032/4275.  
\end{acknowledgements}

\bibliography{ogle182}

\end{document}